\documentclass[final,5p,times,twocolumn,authoryear]{elsarticle}

\usepackage{graphicx}
\usepackage{epstopdf}
\usepackage{dcolumn}
\usepackage{bm}
\usepackage{epsf}
\usepackage{epsfig}
\usepackage{verbatim}
\usepackage{amsmath}
\usepackage{amsfonts}
\usepackage{ifthen}
\usepackage[normalem]{ulem}
\usepackage[flushleft]{threeparttable}
\usepackage{aas_macros}
\usepackage[percent]{overpic}
\usepackage{tabularx}
\usepackage{xcolor}
\definecolor{darkgreen}{rgb}{0.0,0.5,0.0}
\usepackage[colorlinks,citecolor=darkgreen]{hyperref} 

\newcommand{\ie}{\emph{i.e.} }
\newcommand{\eg}{\emph{e.g.,} }

\newcommand{\be}{\begin{equation}}
\newcommand{\ee}{\end{equation}}
\newcommand{\bea}{\begin{equation*}}
\newcommand{\eea}{\end{equation*}}
\newcommand{\beq}{\begin{equation} }
\newcommand{\eeq}{\end{equation}}
\newcommand{\beqr}{\begin{eqnarray} \nonumber}
\newcommand{\eeqr}{\end{eqnarray}}
\newcommand{\beqrb}{\begin{eqnarray}}
\newcommand{\eeqrb}{\nonumber \end{eqnarray}}
\newcommand{\fin}{\mbox{ .}}
\newcommand{\coma}{\mbox{ ,}}

\newcommand{\cm}{\mbox{ cm}}

\newcommand{\se}{\mbox{ s}}

\newcommand{\Myr}{\mbox{ Myr}}

\newcommand{\MHz}{\mbox{ MHz}}

\newcommand{\km}{\mbox{ km}}

\newcommand{\kpc}{\mbox{ kpc}}
\newcommand{\Mpc}{\mbox{ Mpc}}
\newcommand{\eV}{\mbox{ eV}}
\newcommand{\keV}{\mbox{ keV}}

\newcommand{\GeV}{\mbox{ GeV}}
\newcommand{\TeV}{\mbox{ TeV}}

\newcommand{\muG}{\mbox{ $\mu$G}}

\newcommand{\gama}{$\gamma$}

\newcommand{\sh}{{s}}
\newcommand{\myv}{{\upsilon}}
\newcommand{\myK}{{C}}
\newcommand{\gmin}{{\gamma_{1}}}
\newcommand{\gmax}{{\gamma_{2}}}
\newcommand{\gminP}[1]{{\gamma_{1}^{#1}}}
\newcommand{\gmaxP}[1]{{\gamma_{2}^{#1}}}
\newcommand{\nG}{{\mbox{ nG}}}

\begin{document}

\begin{frontmatter}
\title{Macroscopic magnetization of primordial plasma by virial shocks}

\author{Uri Keshet$^1$ and Kuan-Chou Hou$^2$}
\address{
    $^1$ Physics Department, Ben-Gurion University of the Negev, POB 653, Be'er-Sheva 84105, Israel; keshet.uri@gmail.com\\
    $^2$ Institute of Astronomy and Astrophysics, Academia Sinica, PO Box 23-141, Taipei 10617, Taiwan\\
}

\date{\today}

\begin{abstract}
Galaxy-cluster virial (structure-formation accretion) shock observations are shown to imply $\gtrsim1\%$ magnetization of a layer extending $\gtrsim10^{16}$ Debye lengths downstream, challenging the modelling of high Alfv\'en-Mach collisionless shocks.
Unlike similar shocks in supernova remnants or relativistic shocks in $\gamma$-ray burst afterglows, where macroscopic magnetized layers were detected but purportedly attributed to preexisting or non-resonant cosmic-ray streaming-seeded substructure, the upstream of strong virial shocks is both weakly magnetized and pristine.
Hence, some mechanism must generate large-scale and possibly self-similar magnetic sub-structure out of the accreted primordial plasma; such a mechanism may dominate other high-Mach shock systems, too.
\end{abstract}

\end{frontmatter}

\section{Introduction}

Collisionless shocks, central to astrophysics as tracers of fast flows and as sources of cosmic-rays (CRs) and seed magnetic fields, are still not understood from first principles, despite decades of research \cite[for reviews, see][]{TreumannEtAl09Review, BykovTreumann11, SironiEtAl15_review, VanthieghemEtAl20}.
CRs are generally understood to arise from first-order Fermi, so-called diffusive shock acceleration, and several mechanisms for shock magnetization were identified.
However, in the absence of a self-consistent model for the non-linear, multi-scale, three-dimensional shock structure, it is not possible at present to compute macroscopic shock properties of interest, such as the fractional energies deposited in magnetic fields and in different particle species, or the configuration and extent of the shock-magnetized region.
Such parameters are directly needed for modeling the radiative and CR output of the shock, and are indirectly important for understanding the energetics, stability, and various other properties of the host system.

The magnetic-field structure, in particular, determines the synchrotron emission from the shock and its polarization, and affects the subsequent evolution of the plasma and even the energy spectra of CRs and their radiative signature.
Notably, although the CR spectrum is thought to be independent of magnetic-field details for non-relativistic shocks \citep{Krymskii77, AxfordEtAl77, Bell78, BlandfordOstriker78}, provided that the magnetic structure is frozen in the fluid frame and scattering is sufficiently isotropic \citep{keshet2020diffusive}, this is no longer true for relativistic shocks \citep{Kirk_2000, KeshetWaxman05DSA} in systems of more than one dimension \citep{Keshet17DSA1D, lavi2020diffusive}, where anisotropic fields can lead to extreme spectra \citep{AradEtAl19}.
More importantly for the present study, the microscopically vast extents of the magnetized layers inferred from observations downstream of shocks impose a considerable challenge for collisionless shock modelling.

In supernova remnants (SNRs), extended near-equipartition magnetic fields $B$ are inferred downstream of the shock, although the far upstream field $B_u$ is dynamically insignificant, with a typical magnetic-to-kinetic energy fraction $\sigma\equiv M_A^{-2} = B_u^2/(4\pi n_u \bar{m} \myv_u^2)\simeq 10^{-4}$.
Here, $n_u$ and $\myv_u$ are the upstream rest-frame number density and shock velocity, $M_A$ is the Alfv\'enic Mach number, and $\bar{m}$ is the mean particle mass.
In particular, high-resolution X-ray observations imply $B/B_u\simeq 10$--$50$, with such elevated $B$ values persisting $\gtrsim 10^{17}\cm$ downstream
of the shock
\citep{VinkLaming03, BambaEtAl03, VolkEtAl05, ResslerEtAl14, Vink20_Book}.
Hence, strongly magnetized substructure must be seeded near the shock and survive extremely far (in microscopic terms) downstream, over $\gtrsim10^{12}l_D$ Debye lengths or $\gtrsim10^{9}r_L(B)$ Larmor radii.
These scales vastly exceed the shock width and natural scales of thermal plasma instabilities, implying an additional mechanism magnetizing the plasma on large-scales, and the possible emergence of a self-similar microscopic plasma configuration \citep{KatzEtAl07}.

Such macroscopic magnetization cannot arise from electromagnetic, Weibel-like \citep{Fried59} streaming instabilities, which mediate the shock in the non-magnetized limit, without considerable subsequent growth of the magnetic coherence length $l_B$, as unstable Debye-scale modes rapidly decay downstream.
Large-scale magnetization can however be produced if the shock encounters sufficient preexisting, extended upstream density
inhomogeneities, as the resulting Richtmeyer–Meshkov-like rippling of the shock surface can induce sustained downstream turbulence \citep{GiacaloneJokipii07, InoueEtAl09, BeresnyakEtAl09, GuoEtAl12}.
Even in the absence of such preexisting inhomogeneities, nonlinear modes developing in the shock precursor due to non-resonant CR streaming instabilities \citep{Bell04, NiemiecEtAl08, riquelme_09, AmatoBlasi09, StromanEtAl09} may induce macroscopic magnetization in non-oblique shocks \citep{CaprioliSpitkovsky13, CaprioliSpitkovsky14b}, provided that the far upstream is sufficiently magnetized to feed large $l_B$ modes and carry the Alfv\'en waves needed to initiate particle acceleration.
Such instabilities require
$\myv_A\gg \eta \myv_d$ to avoid quenching by filamentation \citep{riquelme_09}, marginally satisfied in SNRs where $B_u\gg 0.4\eta_{3}\muG$.
Here, $\myv_A$ and $\myv_d$ are the Alfv\'en and downstream velocities, and $\eta\equiv 10^{-3}\eta_{3}$ is the upstream number density ratio between CRs and thermal plasma.

An analogous situation is inferred around the relativistic shocks of \gama-ray burst (GRB) afterglows \citep{Gruzinov_Waxman_99, Gruzinov_2001, KatzEtAl07}, in which
$\sigma\simeq 10^{-9}$ and the magnetized layer is again vast in microscopic terms, suggesting the emergence of a self-similar plasma configuration \citep{KatzEtAl07}.
For instance, about a day after the burst, a near-equipartition magnetic field must persist over a downstream layer of proper width $\gtrsim 10^{17}\cm$, corresponding to $\gtrsim 10^{10}$ skin depths $l_{sd}=c/\omega_p$ for typical parameters, where $\omega_p$ is the plasma frequency and $c$ is the speed of light.
Electromagnetic modes on the order of $l_{sd}$ \citep{Medvedev_Loeb_99, Gruzinov_Waxman_99, Silva_etal_2003, nishikawa_05, Spitkovsky_2005, Kato07} rapidly dissipate downstream, so cannot explain such a layer unless they somehow evolve into increasingly large structures \citep{Gruzinov_2001, MedvedevEtAl05, KatzEtAl07, medvedev_zakutnyaya_09, PelletierEtAl09, RevilleBell14, Lemoine_15_Nonlinear}.

Alternatively, as in SNRs, macroscopic magnetic layers could arise downstream of GRB afterglow shocks if sufficient preexisting density inhomogeneities are present far upstream \citep{SironiGoodman07, MilosavljevicNakar07, Goodman_2007}.
Such substructure could also arise from non-resonant CR streaming \citep{MilosavljevicNakar06, reville_06, sironi_spitkovsky_11b} or other \citep{PelletierEtAl09, CasseEtAl13, RevilleBell14, LemoineEtAl14b} instabilities in the precursor of the relativistic shock \citep{MilosavljevicNakar06, reville_06, sironi_spitkovsky_11b}, but only when $\sigma$ is sufficiently large \citep[\eg][]{LemoineEtAl14b}, and if upstream conditions suffice for the shock to produce  magnetization on a range of scales needed to initiate CR acceleration \citep[\eg][]{LemoineEtAl06}.

\begin{figure}[h!]
    \centerline{
    \includegraphics[width=1\linewidth]{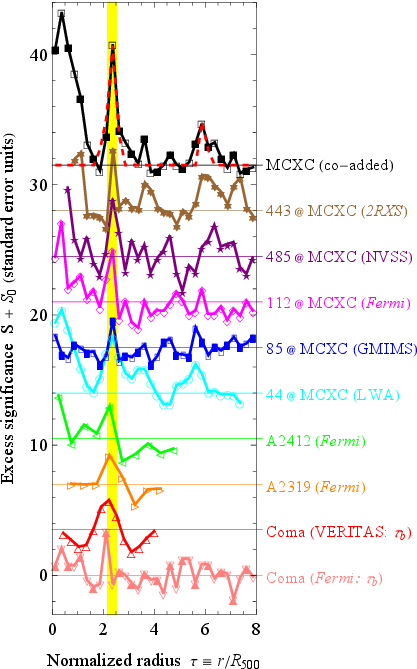}
    }
	\caption{\label{fig:summary}\!\!\!Virial shock signals identified in individual (triangles) or stacked (other symbols) clusters, in diffuse emission (empty symbols), discrete sources (filled), or without separating the two (intermittent empty and filled symbols).
    The significance $S$ (symbols with lines to guide the eye, in standard-error units) of the excess above the background $S_0$ (labelled horizontal lines, shifted vertically for visibility) is plotted as a function of the normalized radius $\tau$, for (bottom to top) clusters Coma, in \emph{Fermi}-LAT \cite[down triangles;][]{keshet2018evidence} and VERITAS \cite[coincident with synchrotron emission and an SZ drop; up triangles;][]{KeshetEtAl17} data as a function of $\tau_b$, A2319 (right triangles) and A2142 (left triangles) in \emph{Fermi}-LAT data \cite[coincident with SZ drops;][]{keshet20coincident}, and for stacked MCXC clusters (labels specify sample sizes) in LWA \cite[circles;][]{HouEtAl23}, polarized GMIMS \cite[rectangles;][]{Keshet24GMIMS}, and \emph{Fermi}-LAT \cite[diamonds;][]{ReissKeshet18} data, and in NVSS (five-stars) and 2XRS (six-stars) source catalogs \citep{IlaniEtAl24a}.
    Also shown are the $2.2<\tau<2.5$ radial range \citep{ReissKeshet18} of MCXC virial-shock signals (vertical yellow band), the co-added MCXC excess (black squares), and a corresponding cylindrical shock model \cite[dashed red curve;][]{Keshet24GMIMS}.
	}
\end{figure}

The observed macroscopic downstream magnetization thus challenges the modelling of both non-relativistic shocks in SNRs, and their relativistic counterparts in GRB afterglows; both systems can be explained either by preexisting upstream density inhomogeneities, or by precursor inhomogeneities generated by combined upstream magnetization and CR streaming.
These explanations can now be tested in strong virial, \ie structure-formation accretion, shocks, which have non-relativistic $\sim10^3\km\se^{-1}$ velocities similar to SNR shocks, but propagate into pristine, primordial infalling gas, which is very poorly magnetized and should not contain any significant substructure.
We find that recent observations of virial shocks around galaxy clusters (see Fig.~\ref{fig:summary}), and in particular their stacked synchrotron emission observed with the Owens Valley  Radio Observatory Long Wavelength Array \cite[OVRO-LWA, hereafter LWA;][]{HouEtAl23} and the Global Magneto-Ionic Medium Survey \cite[GMIMS, radially polarized;][]{Keshet24GMIMS}, indicate downstream magnetization on macroscopic lengthscales, exceeding (in natural units) those inferred from SNRs.

We review observations of virial shocks and their upstream conditions in \S\ref{sec:VS_data}, constrain their downstream magnetization in \S\ref{sec:VS_magnetization}, and conclude with some implications for collisionless shock modelling in \S\ref{sec:Discussion};
a simple virial-shock synchrotron model is outlined in \ref{append:model}.
We adopt a flat $\Lambda$CDM cosmological model with a Hubble constant $H$ of present value $H_0=70\km\se^{-1}\Mpc^{-1}$, an $\Omega_{m}= 0.3$ mass fraction, an $f_b = 0.17$ cosmic baryon fraction, and a $\Gamma=5/3$ adiabatic index for the plasma.
Assuming a $76\%$ hydrogen mass fraction yields a mean $\bar{m} \simeq 0.59 m_p$ particle mass, where $m_p$ is the proton mass.

\section{Observations of virial shocks and their environment}
\label{sec:VS_data}

Faint radiative signatures, predicted from virial shocks \citep{LoebWaxman00, TotaniKitayama00, WaxmanLoeb00, KeshetEtAl03, Miniati02, KeshetEtAl04, KocsisEtAl05}, were identified in data stacked over many low-redshift galaxy clusters \citep{ReissEtAl17, ReissKeshet18, HouEtAl23, Keshet24GMIMS} once lengthscales were normalized by the $R_{500}$ radius of each cluster.
Here, $R_{500}$ encloses a mean density $\rho_c \delta$ of contrast $\delta=500$, where $\rho_c$ is the critical mass density of the Universe.
Such stacked signals, based on the Meta-Catalog of X-ray detected Clusters of galaxies \cite[MCXC;][]{PiffarettiEtAl11} and at least two additional cluster catalogs \citep{IlaniEtAl24,Nadler24InPrep}, were found narrowly confined near the same scaled $\tau\equiv r/R_{500}\simeq 2.4$ cluster radius in very different channels, as demonstrated in Fig.~\ref{fig:summary}.
Similar, coincident signals were found in select nearby clusters \citep{KeshetEtAl17, keshet2018evidence, HurierEtAl19, keshet20coincident}, as illustrated in the figure (triangle symbols).
The figure also shows (stars) the surprising coincident excess of discrete, cataloged radio and X-ray sources, energized by the virial shock \citep{IlaniEtAl24a, IlaniEtAl24}.

Virial shocks are not expected to be spherical. The narrow $\tau\simeq 2.4$ excess likely corresponds to the smallest distance from the center of the cluster (the semi-minor axis in an elliptic approximation), whereas a broad, tentative excess around $\tau\simeq 6$ may correspond to the farthest (semi-major) distance \citep{HouEtAl23, IlaniEtAl24a}, as illustrated in the figure by a simple cylindrical shock model of base radius $\tau=2.4$ and half-height $\tau\simeq6$ \cite[projected, averaged over all orientations, and binned, on top of a fixed background; see][]{Keshet24GMIMS}.
The narrow excess roughly coincides with the dark-matter splashback radius, inferred from a localized drop in the logarithmic radial slopes of galaxy density profiles \citep{MoreEtAl16, ShinEtAl19}.

The diffuse synchrotron, inverse-Compton, and Sunyaev-Zel'dovich (SZ) signatures of the virial shock constrain the electron acceleration $\xi_e\simeq 1\%$ and magnetization $\xi_B\simeq \mbox{few \%}$ efficiencies, the $\gtrsim 10$ Mach numbers of the shocks, and the $\dot{M}/(MH)\simeq 1$ mass accretion rate onto low-redshift clusters \citep{ReissEtAl17, ReissKeshet18, HouEtAl23, keshet2018evidence, HurierEtAl19, keshet20coincident}.
These estimates are based on the diffuse emission remaining after a careful removal of discrete sources, now known \citep{IlaniEtAl24a, IlaniEtAl24} to peak locally at the shock; the residual diffuse radio emission strengthens, as expected \citep{WaxmanLoeb00, KeshetEtAl03}, with cluster mass $M$ \citep{HouEtAl23}.
The significant drop in projected SZ pressure at the shock, the shock strength (inferred from SZ and from the flat spectra of both inverse-Compton and synchrotron signals), the brightness of the leptonic signals, and anecdotal evidence \citep[\eg][]{KeshetEtAl17} all indicate that the $\tau\simeq 2.4$ signals arise from the smooth accretion of primordial, cold gas, and not from heated gas accreted along filaments or streams.

The magnetic field far upstream of virial shocks is highly uncertain.
Faraday rotation imposes model-dependent upper limits on intergalactic-medium (IGM) magnetic fields, such as
$B<4\nG$ fields on $\sim$Mpc scales \citep{OSullivanEtAl20},
$B<30\nG$
in large-scale filaments on $\gtrsim 1\Mpc$ scales \citep{AmaralEtAl21, CarrettiEtAl22, MtchedlidzeEtAl24},
and $B<10\mbox{--}30\nG$ on small, $0.07\mbox{--}0.20\kpc$ scales at redshifts $1\lesssim z\lesssim2$ \citep{PadmanabhanLoeb23}.
Mean $40\mbox{--}80\nG$ fields recently attributed to $z=0$ filaments \citep{CarrettiEtAl23} pertain to warm, virialized regions of much higher densities than upstream of virial shocks, where the gas overdensity $\delta_g\lesssim10$ (with respect to the cosmic mean gas density; see \S\ref{sec:VS_magnetization}) is low.
Indeed, comparable, $30\mbox{--}60\nG$ fields were associated with the claimed detection of synchrotron emission from stacked filaments \citep{VernstromEtAl21}, attributed to shocked regions with amplified $B$ \citep{VernstromEtAl23,Keshet24GMIMS}.
Extrapolating $z=0$ filament estimates to $\delta_g \simeq 10$ suggests $8\mbox{--}26\nG$ \citep{CarrettiEtAl23} fields, but such an extrapolation again pertains at least in part to virialized gas.
We conclude that upstream of the strong virial shocks of galaxy clusters, accreting from voids rather than filaments, $B$ does not exceed a few nG, and could be much weaker.

\section{Virial-shock magnetization}
\label{sec:VS_magnetization}

The spectra of inverse-Compton \citep{keshet2018evidence, ReissKeshet18} and synchrotron \citep{HouEtAl23, Keshet24GMIMS} emission from virial shocks, and the coincident sharp drop in SZ signal \citep{keshet20coincident}, are consistent with the strong shock limit, as expected for the virialization of primordial, cold gas.
The shock velocity is then given by
\begin{equation}
    \myv_u \simeq 4\myv_d \simeq \left(\frac{16k_B T_d}{3 \bar{m} }\right)^{1/2} \simeq 10^3 T_1\km\se^{-1}\coma
\end{equation}
comparable to SNR shock velocities.
Here, $T_d\equiv T_1k_B^{-1}\keV$ is the downstream temperature, subscripts $u$ ($d$) denote upstream (downstream) quantities, and $k_B$ is the Boltzmann constant.

For simplicity, we model the gas distribution using an isothermal $\beta$-model with the typical $\beta=2/3$ of galaxy clusters, but take into account that due to peripheral steepening, the gas density may diminish by an additional factor $f_\beta\simeq 1/3$ by the virial-shock radius \citep{HouEtAl23}.
The particle number density upstream of the virial shock is then given by \cite[see \ref{append:model} and][]{HouEtAl23}
\begin{equation}\label{eq:nu}
  n_u \simeq \frac{n_d}{4}\simeq \frac{f_\beta f_b \delta_s \rho_c}{12\bar{m}} \simeq 10^{-5}f_\beta\,\delta_{90}\cm^{-3}\coma
\end{equation}
where subscript $s$ pertains to the shock radius,
\begin{equation}\label{eq:delta_s}
  \delta_s \simeq 500\tau_s^{-2} \simeq 87 (\tau_s/2.4)^{-2} \coma
\end{equation}
and $\delta_{90}\equiv \delta_s/90$.

The upstream gas density \eqref{eq:nu} is higher than its cosmic mean value by a factor
\begin{equation}\label{eq:fg}
  f_g = \frac{\bar{m}n_u}{f_b\Omega_m \rho_c}
  \simeq \frac{f_\beta \delta_s }{12\Omega_m} \simeq 25 f_\beta\,\delta_{90}\coma
\end{equation}
so pre-shock compression can amplify primordial magnetic fields $B_p$ by a small factor $B_u/B_p\simeq f_g^{2/3}$ (in the isotropic case).
The IGM magnetic field estimates reviewed in \S\ref{sec:VS_data} indicate that unlike SNR shocks, the upstream of virial shocks is weakly magnetized, with plasma beta
\begin{equation}\label{eq:PlasmaBeta}
  \beta_p \simeq \frac{n_u k_B T_u}{B_u^2/8\pi}\simeq 2400 B_1^{-2} f_\beta T_{10}\delta_{90} \coma
\end{equation}
where $T_{10}\equiv k_B T_u/10\eV\simeq 1$ according to cosmological simulations \citep{KeshetEtAl03}, and $B_1\equiv B_u/1\nG$.
Such upstream fields are far less significant dynamically than in SNR shocks,
\begin{equation}\label{eq:UpstreamSig}
  \sigma \equiv \frac{B_u^2}{4\pi n_u \bar{m} \myv_u^2} \simeq 8\times 10^{-7} \frac{B_1^{2}}{f_\beta T_1 \delta_{90}} \fin
\end{equation}

The properties of the downstream magnetic field can be inferred from the observed radio signal.
In the strong-shock limit, cosmic-ray electron (CRE) injection with a flat energy spectrum of index $p=2$ yields a steady-state volume-integrated (dimensionless) distribution
\begin{equation}\label{eq:CRESpectrum}
\frac{dN_e}{d\gamma} \simeq C \gamma^{-3}
\end{equation}
in a cooling-limited range of Lorentz factors $\gamma<\gamma_{max}$, where $C$ is a dimensionless constant.
These CREs radiate a spectrally flat synchrotron luminosity
\begin{equation}
\nu L_\nu \simeq \frac{c \sigma_T C}{12\pi}B^2 \coma
\label{eq:symRLp2}
\end{equation}
where $\sigma_T$ is the Thomson cross section and $\nu$ is the frequency; estimates of the projected flux and brightness profiles are provided by \cite{HouEtAl23}.
Here and below, $B^2$ is defined as the mean squared magnetic-field amplitude in the region traversed by the radiating CREs before they cool.
The magnetic energy fraction
\begin{equation}\label{eq:xiB}
  \xi_B\equiv \frac{B^2/8\pi}{(3/2)n_dk_B T_d}
\end{equation}
was estimated as $\sim 1\%$ by modeling the radio signal from Coma \citep{KeshetEtAl17},
whereas the stacked LWA excess implies a somewhat stronger magnetization in the range $\xi_B\simeq (2\mbox{--}9)\%$ \citep{HouEtAl23}.

CREs at cluster peripheries cool primarily by inverse-Compton scattering cosmic-microwave background (CMB) photons, with a characteristic cooling time
\begin{equation}\label{eq:tcool}
  t_{\mbox{\scriptsize{cool}}} \simeq \frac{3m_e c}{4u_{cmb}\sigma_T \gamma}
    \simeq 280 \,\nu_{76}^{-1/2}(T_1 f_\beta\delta_{90}\xi_5)^{1/4}\Myr \coma
\end{equation}
where the last estimate focuses on CREs synchrotron-radiating in LWA frequencies, $\nu_{76}\equiv \nu/76\MHz\simeq 1$.
Here, $u_{cmb}$ is the CMB energy density, $m_e$ is the electron mass, and we defined $\xi_5\equiv \xi_B/0.05$.
At the highest CRE energies, the acceleration time $t_{acc}\simeq r_L c/(4\myv_d)^2$ is $t_0\equiv t_{acc}(\gamma_{max})=t_{\mbox{\scriptsize{cool}}}(\gamma_{max})$, yielding
\begin{equation}\label{eq:gmax}
  \gamma_{max}
  \simeq \frac{\myv_u}{c} \left(\frac{3eB}{4u_{cmb}\sigma_T}\right)^{1/2}
  \simeq 7\times 10^7T_1^{3/4}(f_\beta\delta_{90}\xi_5)^{1/4} \coma
\end{equation}
where $e$ is the electron charge and $r_L$ is the Larmor radius.
This upper limit lies a factor of a few above the $\sim9\TeV$ energy of CREs radiating in $\sim220\GeV$ VERITAS energies \citep{KeshetEtAl17} and in the high-energy band of the stacked \emph{Fermi}-LAT signal \citep{ReissKeshet18}.

The acceleration of CREs to high energies requires strong magnetization around the shock, over a wide range of scales, needed to facilitate the Fermi cycles.
The observed virial-shock \gama-ray signals thus impose a lower limit on the width $W$ of the downstream magnetized layer,
\begin{equation}\label{eq:AccelerationLayer}
  W>\frac{r_L c}{\myv_d}
  \simeq \frac{\gamma_{max} m_e c^3}{e B \myv_d}
  \simeq \frac{0.13}{(f_\beta T_1 \delta_{90}\xi_5)^{1/4}} \kpc \fin
\end{equation}
In addition, a $\myv_u/\myv_d\simeq 4$ times narrower magnetized precursor is necessary upstream.
Such magnetized layers are insufficiently wide, however, to account for the strength of the observed synchrotron signals \citep{KeshetEtAl17, HouEtAl23, Keshet24GMIMS}.
The low energy, $\gamma\ll\gamma_{max}$ CREs radiating at such radio frequencies escape the shock much faster than at high energies, after only a $\sim t_0\gamma/\gamma_{max}$ duration, and cool over a much longer, $t_0\gamma_{max}/\gamma$ time.
As $B^2$ pertains to the field traversed during $t_{\mbox{\scriptsize{cool}}}$, and the corresponding $\xi_B$ is already close to equipartition, much of the radio signal is emitted while these CREs are advected away from the shock, requiring an extended downstream magnetized region of width
\begin{equation}\label{eq:WidthAdvection}
  W > \myv_d t_{\mbox{\scriptsize{cool}}}
   \simeq  66 T_1^{3/4}\nu_{76}^{-1/2}(f_\beta \delta_{90}\xi_5)^{1/4} \kpc \fin
\end{equation}
In addition to advection, these CREs also diffuse as they radiate, but only over a relatively short distance $(r_L c t_{\mbox{\scriptsize{cool}}})^{1/2}\ll \myv_d t_{\mbox{\scriptsize{cool}}}$.

We conclude that accounting for the observed radio emission, and in particular the low-frequency LWA signal, requires an extended, macroscopic magnetized layer downstream of virial shocks.
This layer is considerably thicker than required in SNR shocks, not only in its proper width, but also in terms of relevant physical quantities.
In units of the downstream proton skin depth,
\begin{equation}\label{eq:lsd}
  l_p \equiv \left(\frac{4\pi n_d e^2}{m_p}\right)^{1/2} \simeq \frac{3.3\times 10^4}{(f_\beta\delta_{100})^{1/2}}\km \coma
\end{equation}
the downstream magnetized layer has a minimal width
\begin{equation}\label{eq:AccelerationLayerLp}
  W/l_p >
  6\times 10^{13}\left(T_1 f_\beta \delta_{90}\right)^{3/4}\nu_{76}^{-1/2}\xi_5^{1/4} \fin
\end{equation}
More importantly, in terms of the downstream Debye length,
\begin{equation}\label{eq:Debye}
  l_D \simeq \left(\frac{k_B T_d}{4\pi n_d e^2}\right)^{1/2} \simeq 34 \left(\frac{T_1}{f_\beta\delta_{90}}\right)^{1/2}\km \coma
\end{equation}
the width becomes
\begin{equation}\label{eq:AccelerationLayerLD}
  W/l_D >  6\times10^{16}\left(f_\beta \delta_{90}\right)^{3/4}\nu_{76}^{-1/2}\left(T_1\xi_5\right)^{1/4} \fin
\end{equation}
Similarly, $W/r_L(B)>5\times 10^9T_1^{3/2}f_\beta \delta_{90}\xi_5/\nu_{76}$, but the measured $B^2$ does not provide a reliable estimate of the Larmor radius in the patchy field expected downstream.

\section{Discussion}
\label{sec:Discussion}

The extended, $W\gtrsim 10^{16}l_D$ magnetized layers we infer downstream of virial shocks challenge the modeling of such weakly-magnetized, \ie high $M_A$, shocks; this challenge is qualitatively more difficult than in the cases of SNR and GRB afterglow shocks.
Attributing macroscopic magnetization to sufficient pre-existing density inhomogeneities is rather artificial even for SNR shocks, but becomes quite implausible for strong virial shocks, which propagate into pristine, non-virialized gas.
Density inhomogeneities from non-resonant CR streaming instabilities in the shock precursor are also less likely in virial shocks, considering their weak upstream magnetization.
In the weak fields of Eq.~\eqref{eq:PlasmaBeta}, the onset of CR acceleration may be greatly delayed, until sufficient magnetic substructure is generated
to sustain the Fermi cycles.
Non-resonant CR streaming instabilities must then compete with filamentation modes, as $\sigma\ll 10^{-3}$ by Eq.~\eqref{eq:UpstreamSig}, and may become entirely quenched if the (poorly-constrained) upstream fields are sufficiently weak.
For instance, the $\myv_A\gg\eta\myv_d$ growth condition
may be violated in virial shocks, as it requires
\begin{equation}\label{eq:NoStreaming}
  B_u \gg 0.3\eta_3(f_\beta T_1 \delta_{90})^{1/2}\nG \fin
\end{equation}

If non-resonant CR streaming instabilities are indeed quenched or greatly diminished in virial shocks, then the macroscopic magnetization we infer requires an alternative origin, such as nonlinear Weibel-like modes evolving hand-in-hand with particle acceleration to generate large-scale substructure, possibly developing a self-similar configuration \citep{KatzEtAl07}.
Fully ab-initio particle-in-cell (PIC) simulations have so far resolved only the initial stages of such shock evolution, mainly in 2D relativistic pair-plasma flows \citep{spitkovsky2008particle, KeshetEtAl09PIC, LemoineEtAl19A, GroseljEtAl24}.
Early modes are limited to small-scales and decay too quickly \citep{Gruzinov_2001}, as $k^3c^3/\omega_p$ \citep{ChangEtAl08, Lemoine_15_Nonlinear}, but the gradual acceleration of particles \citep{spitkovsky2008particle} drives magnetic substructure on increasingly long wavelengths $\lambda$, surviving farther downstream with a decay rate \citep{KeshetEtAl09PIC}
\begin{equation}\label{eq:Decay}
  \Gamma_B \equiv -\frac{d\ln B}{dt} \simeq  0.7 \left(\frac{\lambda}{l_{sd}}\right)^{-2}\omega_p \fin
\end{equation}
The slow $l_B$ growth during the accessible, $t\sim[10^3,10^4]\omega_p^{-1}$ downstream simulation times
is associated with nonlinear, solitary low-density modes (cavities), emerging upstream following nonlinear filamentation \citep{KeshetEtAl09PIC, GroseljEtAl24}, and resulting in $\sim10\%$ equipartition fields in $1\%$ of the downstream area \citep{GroseljEtAl24}.
Note that although the impossibility of a 2D dynamo and other limitations \citep[\eg][]{Gruzinov08} do not rule out such kinetic upstream evolution, the 3D case is likely very different; a realistic ion-to-electron mass should also modify the picture.

If such a mechanism does dominate virial shocks, then it may well operate at some level also in other low $\sigma$ systems, such as
SNR and GRB afterglow shocks.
Hybrid simulations \citep{CaprioliSpitkovsky13, CaprioliSpitkovsky14b} have established the dominance of non-resonant CR-streaming instabilities in non-oblique, non-relativistic shocks in the range $30\lesssim M_A\lesssim 100$ for interstellar-medium (ISM) magnetization levels, but it is unclear if this mechanism remains dominant when taking into account kinetic electron effects over realistic length and time scales, and under more general physical conditions such as weaker magnetization.

\paragraph*{Acknowledgements} \begin{small}
We thank Y. Lyubarsky, I. Demidov, E. Waxman, and M. Gedalin for insightful discussions.
This research was supported by the Israel Science Foundation (ISF grant No. 2126/22).
\end{small}

\bibliography{Virial}

\appendix

\section{Virial-shock synchrotron model}
\label{append:model}

We reproduce the virial shock model of \cite{HouEtAl23}, including its synchrotron estimate.
We begin with the isothermal $\beta$-model, in which the particle number density follows
\begin{align}
    n(r) = n_{0} \left[ 1 + \left(\frac{r}{r_c}\right)^2 \right]^{-3\beta/2} \coma
    \label{eq:beta_model}
\end{align}
where $n_{0}$ is the central number density, $r_c$ is the core radius, and $\beta$ is the slope parameter.
In the hydrostatic equilibrium limit, the total (gravitating) mass inside a radius $r$ is given by
\begin{equation}
    M(r) \simeq \frac{3 \beta k_B T r}{G \bar{m}} \left(1 + \frac{r_c^2}{r^2} \right)^{-1} \coma
    \label{eq:hydro_eq_mass}
\end{equation}
where $G$ is Newton’s constant.
The typical $r_c \sim 0.1 R_{500}$ core radius is much smaller than the $r\gtrsim 2 R_{500}$ radii of interest, so we may approximate the downstream distribution as $n(r) \simeq n_{0} (r/r_c)^{-3 \beta}$, and the corresponding mass as
\begin{equation}
    M(r) \simeq \frac{3 \beta k_B T}{G \bar{m}}r \fin
    \label{eq:hydro_eq_appr_mass}
\end{equation}

In this approximation, the over-density parameter $\delta=500/\tau^2$ yields $\delta_s(\tau_s\simeq 2.4)\simeq 90$,
where subscript $\sh$ denotes evaluation at the shock radius.
The enclosing radius
\begin{equation}
    R_\delta = \left[\frac{9 \beta k_B T}{4 \pi \rho_c(z) \delta G \bar{m}}\right]^{1/2}
    \label{eq:hydro_eq_appr_mass}
\end{equation}
and mass-temperature relation
\begin{align} \label{eq:MassTemperatureRelation}
M_\delta = \frac{9}{2 \sqrt{\pi \rho_c(z) \delta}} \left( \frac{\beta k_B T}{G \bar{m}}\right)^{3/2} \coma
\end{align}
are defined such the mean mass density  $M_\delta / [ (4/3) \pi R_\delta^3 ]$ enclosed within a radius $R_\delta$ is higher by a factor $\delta$ than the $\rho_c(z) \equiv \rho_0 \mathcal{H}^2$ critical mass density of the Universe, with $\rho_0 = 3 H_0^2 / (8 \pi G)$ being the present value $\rho_c(0)$ and $\mathcal{H} \equiv H(z)/H_0 \simeq [(1-\Omega_m) + (1+z)^3\Omega_m]^{1/2}$ describing the evolution of the Hubble constant.
Assuming that the total baryon mass $f_b M_{\sh}$ is given by the spatial integral of the isothermal $\beta$-profile (\ref{eq:beta_model}) now yields the downstream particle number density,
\begin{equation}
    n_d = (1-\beta)f_\beta \frac{f_b \rho_c(z)}{\bar{m}} \delta_{\sh} \coma
    \label{eq:down_par_density}
\end{equation}
where the factor $f_\beta\simeq 1/3$ was introduced to account for the deviation from the $\beta$-model profile at the cluster periphery.

We assume that a fraction $\xi_e$ of the thermal energy density downstream of the shock is deposited in the CRE distribution \eqref{eq:CRESpectrum}, spanning the $\gmin<\gamma<\gmax$ range.
The effective value of $\gmin$ has little effect on our results, so is taken for simplicity as unity.
The volume-integrated normalization in Eq.~(\ref{eq:CRESpectrum}) is then given by
\begin{equation} \label{eq:myK}
    \myK=\frac{9 k_B T \xi_e}{8(p-1)c u_{cmb} \sigma_T}\frac{f_b \dot{M_{\sh}}}{\bar{m}}\times
    \begin{cases}
    \frac{1}{\ln(\gmax/\gmin)} & \mbox{for $p=2$\,;} \\ \\
    \frac{p-2}{\gminP{2-p} - \gmaxP{2-p}} & \mbox{for $p \neq 2$\,,}
    \end{cases}
\end{equation}
where we adopted the typical $\beta=2/3$ (henceforth).
To estimate the number accretion rate $f_b \dot{M_{\sh}} / \bar{m}$ of gas particles through the shock,
we parameterize the accretion rate as proportional to $M_{\sh}$ by introducing the dimensionless accretion parameter
\begin{equation}
    \dot{m} \equiv \frac{\dot{M_{\sh}}}{M_{\sh} H(z) }\fin
    \label{eq:accretion_rate}
\end{equation}

The synchrotron luminosity radiated by a CRE population of steady-state spectral index $p+1$ is \citep[\eg][]{RybickiLightman86}
\begin{equation}
L_\nu = \alpha_e h\nu\,\myK \Phi q(p)\left( \frac{3\nu_B}{\nu}\right)^{1+\frac{p}{2}}  \coma
\label{eq:symRL}
\end{equation}
where $\alpha_e=e^2/(\hbar c)$ is the fine-structure constant, $h=2\pi\hbar$ is Planck's constant,
$\nu_B=e B/(2\pi m_e c)$ is the cyclotron frequency,
$\Gamma(y)$ is the gamma function,
and the $p$-dependent numerical factor
\begin{equation}
q(p)\equiv \frac{ \,\Gamma\left( \frac{p}{4} + \frac{11}{6}\right) \Gamma\left( \frac{p}{4} + \frac{1}{6}\right)}{(p+2)\sqrt{3}}\,.
\end{equation}
The dependence $\Phi \equiv \sin^{(p+2)/2}{\phi}$ upon the pitch angle $\phi$
is isotropically averaged to give
\begin{equation}
 \langle\Phi\rangle = \frac{\sqrt{\pi}\,\Gamma\left(\frac{p}{4}+\frac{3}{2}\right)}{2 \Gamma\left(\frac{p}{4}+2\right)}  \coma
\label{eq:pitch_angle}
\end{equation}
leading to Eq.~\eqref{eq:symRLp2}.
For the corresponding flux and brightness profiles, see \cite{HouEtAl23}.

\end{document}